# Optimization Design and Analysis of Systematic LT codes over AWGN Channel

Shengkai Xu, Dazhuan Xu, Xiaofei Zhang and Hanqin Shao

College of Electronic and Information Engineering

Nanjing University of Aeronautics and Astronautics

Email: xudazhuan@nuaa.edu.cn

*Abstract*—In this paper, we study systematic Luby Transform (SLT) codes over additive white Gaussian noise (AWGN) channel. We introduce the encoding scheme of SLT codes and give the bipartite graph for iterative belief propagation (BP) decoding algorithm. Similar to low-density parity-check codes, Gaussian approximation (GA) is applied to yield asymptotic performance of SLT codes. Recent work about SLT codes has been focused on providing better encoding and decoding algorithms and design of degree distributions. In our work, we propose a novel linear programming method to optimize the degree distribution. Simulation results show that the proposed distributions can provide better bit-error-ratio (BER) performance. Moreover, we analyze the lower bound of SLT codes and offer closed form expressions.

I. INTRODUCTION

Fountain codes [1, 2] were first introduced over binary erasure channels (BEC). Due to packets loss, such as in Internet, automatic repeat request (ARQ) was adopted. However, request and re-transmission occur frequently in ARQ and that leads to low efficiency. This problem does not exist in Luby Transform (LT) codes. LT codes [3] are the first class of fountain codes to be realized. Given some input symbols, infinite output symbols are generated and then transmitted to the receiver till all the input symbols are recovered. Thus by using LT codes, we do not necessarily need to know the channel state information (CSI). However, decoding complexity of LT codes increases rapidly as the length of input symbols increases, which is not good for practical applications. Raptor codes [4] fix the problem because the decoding complexity of Raptor codes is linearly dependent on the length of input symbols.

The idea of fountain codes over BEC was extended to other channels, for instance, binary symmetric channels (BSC), additive white Gaussian noise (AWGN) channels, and fading channels. The performance of fountain codes on these channels is also satisfying [5-7]. Low-density parity-check (LDPC) codes [8] were first invented by Gallager in 1962. Belief propagation (BP) decoding algorithm was used to decode LDPC codes on noisy channels and it is based on the bipartite graph of parity-check matrix. As LT codes and LDPC codes both have low-density generator matrix, BP algorithm is still suitable for LT codes [9]. Even though LT codes can deal with different channel situations, they still suffer error floor on noisy channels [10].

Systematic codes are popular in many practical applications; however LT codes and Raptor codes are not designed systematically. Taking BEC for an example, if packets are transmitted through the channel and there is few or even no loss, systematic codes would performance way better than nonsystematic codes considering the cost. Several works have been done on systematic

Luby Transform (SLT) codes [11]. Yuan [12] proposed a family of systematic rateless codes for BEC. In [13], Nguyen provided a designed distribution for systematic LT codes. Chen in [14] proposed another design of degree distribution. Zhang [15] gave a new soft decoding method to improve performances. In [16], Hayajneh provided a new encoding algorithm by shaping the left degree distribution away from Poisson distribution. In [17], Asteris introduced a new family of fountain codes that are systematic and have sparse parities. Besides, Chen [18] studied systematic Raptor codes with efficient encoding method.

In this paper, we focus on SLT codes over AWGN channel and analyze the asymptotic performance using Gaussian approximation. More importantly, we give a novel linear programming method to optimize check node degree distribution. Our optimal distributions can provide better performance than the designed distribution in [13]. Last but not least, we derive some lower bound expressions of SLT codes. The remainders of the paper are organized as follows. Section II will introduce the encoding and decoding algorithms of SLT codes. In section III we will use Gaussian approximation to analyze the asymptotic performance and a novel optimization of degree distribution is proposed. Several lower bound expressions are derived for the first time in Section IV and some simulation results are shown in Section V. Section VI will conclude the paper.

## II. SYSTEM MODEL

We consider SLT codes transmitted over AWGN channel. The noise is Gaussian $\mathcal{N}(0, \sigma_n^2)$. One input symbol can be just one bit or a block of bits. Either way, it has no impact on the analysis of SLT codes. So we use a bit to represent an input symbol in this paper. After SLT encoding, we adopt binary phase shift keying (BPSK) modulation. By **c** we denote the encoded

SLT codes and $\mathbf{x} = 1 - 2\mathbf{c}$ the BPSK modulated symbols transmitted to the receiver. By adding channel noise $\mathbf{n}$, the received symbols are $\mathbf{y} = \mathbf{x} + \mathbf{n}$.

*A. Generation of SLT codes*

Let $K$ denote the length of input symbols and $\Omega_1, \Omega_2, \ldots, \Omega_D$ be the degree distribution on set $\{1, 2, \ldots, D\}$ so that $\Omega_i$ denotes the probability that degree $i$ is chosen. Generally we denote such distribution by its generator polynomial, i.e., $\Omega(x) = \sum_{i=1}^{D} \Omega_i x^i$ where $D$ is the maximum degree. The generation of LT codes takes several steps as follows [3].

- Sample a degree $i$ with probability $\Omega_i$ in $\Omega(x)$;
- Sample $i$ different input symbols uniformly at random from $K$ in total and then XOR them;
- Repeat above two steps to get output LT encoded symbols.

LT codes are designed to be rateless but in actual applications the output symbols are finite. Suppose we have $M$ encoded output symbols and let $\varepsilon = M/K - 1$ denote extra percentage of input symbols that are needed at the receiver, which is called overhead. In SLT codes, input symbols are transmitted along with LT encoded symbols. Let $\mathbf{u} = (u_1, u_2, \ldots u_K)$ denote $K$ input symbols. By the generation of LT codes, we can get $M$ encoded symbols, i.e., $\mathbf{c}_{LT} = (c_1, c_2, \ldots, c_M) = \mathbf{u} \cdot G_{LT}$ where $G_{LT}$ is the generator matrix of size $K \times M$ and thus SLT codes are $\mathbf{c}_{SLT} = (\mathbf{u}, \mathbf{c}_{LT})$. By adding a unity matrix of size $K \times K$ into $G_{LT}$ we can get the generator matrix of SLT codes $G = [I \ G_{LT}]$. Let $N = K + M$ denote the length of SLT codes. Similar to LT codes, we define overhead as $\varepsilon = N/K - 1 = M/K$. Afterwards, SLT codes are modulated as BPSK symbols and transmitted to the receiver through AWGN channel.

*B. BP deoding algorithm for SLT codes*

Similar to LDPC codes, decoding SLT codes over AWGN channel is based on the bipartite graph. By $H$ we denote the parity-check matrix of SLT codes as $H = [G_{LT}^T \ I]$ of size $M \times N$, where $G_{LT}^T$ is the transposition of $G_{LT}$. Each row and column in $H$ represents a check node and a variable node, respectively. Fig. 1 gives an instance of bipartite graph based on parity-check matrix of SLT codes, where variable nodes and check nodes are on opposite sides. As can be seen, variable nodes can be divided into two different types, namely, source nodes and check nodes. The log-domain of BP decoding algorithm uses log-likelihood ratios (LLRs), which is

$$\text{LLR}(c_k) = \ln \frac{\Pr(c_k = 0 \mid y_k)}{\Pr(c_k = 1 \mid y_k)} \quad (1)$$

where $y_k$ is the received symbol of the encoded symbol $c_k$. By **Z** we denote the initial LLRs related to channel

$$\mathbf{Z} = \frac{2\mathbf{y}}{\sigma_n^2} \quad (2)$$

In the following, we denote by $R_{m,n}$ and $Q_{n,m}$ the message passing from $n$-th check node to $m$-th variable node and the message passing from $m$-th variable node to $n$-th check node, respectively. By $\{S_n\} \setminus m$ we denote the set of all nodes adjacent to node $n$ except $m$. In round 0 of BP decoding, we initialize variable nodes with messages **Z**. In round $l$, messages passing through bipartite graph [11] will be updated as

$$R_{m,n}^{(l)} = \begin{cases} 2\tanh^{-1}\left[\tanh\left(\frac{Z_{K+n}}{2}\right) \prod_{k \in \{S_n\} \setminus m} \tanh\left(\frac{Q_{n,k}^{(l-1)}}{2}\right)\right], m \leq K \\ 2\tanh^{-1}\left[\prod_{k \in \{S_n\} \setminus m} \tanh\left(\frac{Q_{n,k}^{(l-1)}}{2}\right)\right], m > K \end{cases} \quad (3)$$

$$Q_{n,m}^{(l+1)} = \begin{cases} Z_m + \sum_{k \in \{S_m\} \setminus n} R_{m,k}^{(l)}, m \le K \\ Z_m, m > K \end{cases} \quad (4)$$

Since we only focus on source nodes and it is implied from (3) and (4) that messages between source nodes and check nodes on the same side of bipartite graph are irrelevant and besides messages of those check nodes are unchangeable, we may adjust (3) and (4) to simple ones which we are more interested in, i.e., messages exchanging between source nodes and check nodes on opposite sides of bipartite graph, as in (5) and (6). The corresponding bipartite graph is modified and shown in Fig. 2. From this point on, further discussions will all be based on the new graph.

$$R_{m,n}^{(l)} = 2\tanh^{-1}\left[\tanh\left(\frac{Z_{K+n}}{2}\right) \prod_{k \in \{S_n\} \setminus m} \tanh\left(\frac{Q_{n,k}^{(l-1)}}{2}\right)\right] \quad (5)$$

$$Q_{n,m}^{(l+1)} = Z_m + \sum_{k \in \{S_m\} \setminus n} R_{m,k}^{(l)} \quad (6)$$

During the decoding iterations in the graph of Fig. 2 we may focus on the degree distribution with respect to edges rather than nodes. We denote by $\lambda_i$ the fraction of edges connected to source nodes of degree $i$. And $\Lambda_i$ denotes the probability a source node chosen in the bipartite graph is of degree $i$. We denote by $\lambda(x)$ and $\Lambda(x)$ the generator polynomial $\sum_{i=1}^{d_s} \lambda_i x^{i-1}$ and $\sum_{i=1}^{d_s} \Lambda_i x^i$, respectively, where the maximum degree of source nodes is $d_s$.

Let $\omega(x) = \sum_{j=1}^{d_c} \omega_j x^{j-1}$ denote edge degree distribution of check node, where $\omega_j$ is the fraction of edges connected to check nodes of degree $j$ and $d_c$ is the maximum degree of check nodes. Recall that $\Omega(x)$ is the check node degree distribution. Then we have

$$\lambda(x) = \frac{\Lambda'(x)}{\Lambda'(1)} \tag{7}$$

$$\omega(x) = \frac{\Omega'(x)}{\Omega'(1)} \tag{8}$$

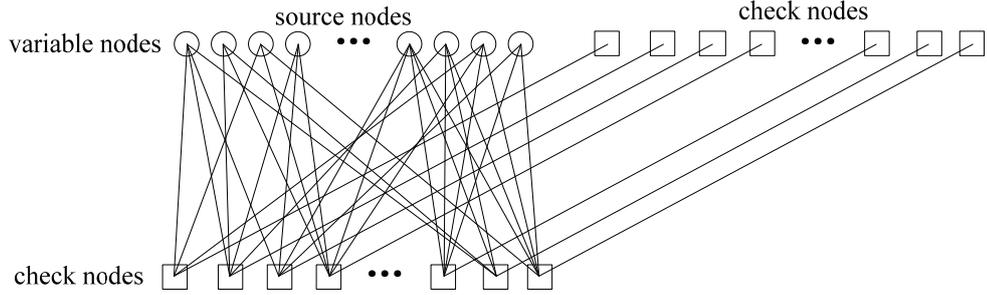

Fig. 1. Bipartite graph of SLT codes, with variable nodes and check nodes on each side.

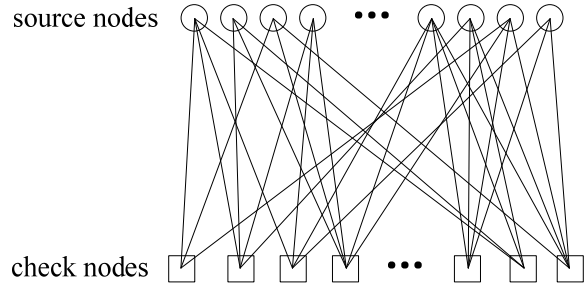

Fig. 2. New bipartite graph of SLT codes, with source nodes and check nodes on each side.

where $f'(x)$ denotes the formal derivative of $f(x)$ with respect to $x$. $\omega(x)$ and $\Omega(x)$ are obviously independent of number of check nodes while $\lambda(x)$ and $\Lambda(x)$ may depend on the number of check nodes. However when $M$ is large enough, $\Lambda(x)$ is considered as Poisson distribution. We denote by $\alpha$ the average degrees of source nodes thus $\Lambda(x) = e^{\alpha(x-1)}$ and $\lambda(x)$ is approximately Poisson distribution as well.

III. ASYMPTOTIC ANALYSIS AND OPTIMIZATION OF SLT CODES

In this section, we use Gaussian approximation to study the asymptotic performance of SLT codes over AWGN channel. Based on BP iterative decoding algorithm, closed form expression of BER is derived. By enforcing the density evolution formulae, we propose a novel linear programming method to optimize degree distribution of SLT codes.

*A. Gaussian approximation(GA)*

Wiberg [19] observed that LLR message distributions for AWGN channels resemble Gaussians for LDPC codes. Due to the similarities of SLT codes and LDPC codes, we assume $R$ and $Q$ in (5) and (6) can be well approximated by Gaussian densities over AWGN channel. Since a Gaussian variable is totally determined by its mean and variance, we only need to keep an eye on means and variances during BP iterations. There is an important symmetric condition $f(x) = f(-x)e^x$ which is preserved under density evolution for all messages, where $f(x)$ is an LLR message density [20]. For approximate Gaussian density with mean $\mu$ and variance $\sigma^2$, by enforcing this symmetric condition we can get $\sigma^2 = 2\mu$. In such case we only need means of Gaussian variables during iterations. Meanwhile we transmit all-zero codes over AWGN channel thus $Z$ is Gaussian $\mathcal{N}(2/\sigma_n^2, 4/\sigma_n^2)$.

Taking expectations of both sides in (5), we get

$$E\left[\tanh\frac{R^{(l)}}{2}\right] = E\left[\tanh\left(\frac{Z}{2}\right)\right] E\left[\tanh\left(\frac{Q^{(l-1)}}{2}\right)\right]^{j-1} \quad (9)$$

where we have omitted the indices because they are i.i.d. Gaussian variables. Product in (5) has also been simplified for the same reason given the check node has $j$ adjacent source nodes for $1 \leq j \leq d_c$. Recall that both $R^{(l)}$ and $Q^{(l-1)}$ are Gaussian variables, i.e., $\mathcal{N}(\mu_R^{(l)}, 2\mu_R^{(l)})$ and

$\mathcal{N}\left(\mu_Q^{(l-1)}, 2\mu_Q^{(l-1)}\right)$, respectively. Apparently, $\mu_R^{(l)} = \sum_{j=1}^{d_c} \omega_j \mu_{R,j}^{(l)}$ and $\mu_Q^{(l-1)} = \sum_{i=1}^{d_s} \lambda_i \mu_{Q,i}^{(l-1)}$, where $\mu_{R,j}^{(l)}$ and $\mu_{Q,i}^{(l-1)}$ are means of message distributions of check nodes with degree $j$ and source nodes with degree $i$, respectively.

Since

$$E\left[\tanh\frac{R}{2}\right] = \frac{1}{\sqrt{4\pi\mu_R}} \int_{-\infty}^{+\infty} \tanh\left(\frac{r}{2}\right) e^{-\frac{(r-\mu_R)^2}{4\mu_R}} dr,$$

we define a function $\varphi(x)$ [21] as in (10) for all $x \in [0, +\infty)$

$$\varphi(x) = \begin{cases} 1 - \frac{1}{\sqrt{4\pi x}} \int_{-\infty}^{+\infty} \tanh\left(\frac{r}{2}\right) e^{-\frac{(r-x)^2}{4x}} dr, & x > 0 \\ 1, & x = 0 \end{cases} \quad (10)$$

Thus, (9) can be rewritten as

$$\mu_{R,j}^{(l)} = \varphi^{-1}\left(1 - \left(1 - \varphi\left(\frac{2}{\sigma_n^2}\right)\right)\left(1 - \sum_{i=1}^{d_s} \lambda_i \varphi\left(\mu_{Q,i}^{(l-1)}\right)\right)^{j-1}\right) \quad (11)$$

where $\varphi^{-1}(x)$ is the inverse function of $\varphi(x)$ and because there is no exist of closed form expression for $\varphi(x)$, we normally use an approximate expression for all $x > 0$

$$\varphi(x) = \begin{cases} e^{-\left(0.4527 x^{0.86} + 0.0218\right)}, & 0 < x < 10 \\ \sqrt{\frac{\pi}{x}}\left(1 - \frac{10}{7x}\right) e^{-\frac{x}{4}}, & x \geq 10 \end{cases} \quad (12)$$

Similarly, taking expectations of (6) we get

$$\mu_{Q,i}^{(l+1)} = \frac{2}{\sigma_n^2} + (i-1)\mu_R^{(l)} \quad (13)$$

After $l$ iterations, the mean of LLR messages for decision is $\mu_i^{(l)} = 2/\sigma_n^2 + i\mu_R^{(l)}$ and the asymptotic performance of BP decoding after $l$ iterations as all-zero codeword transmitted is

$$\begin{aligned}
P_e^{(l)} &= \sum_{i=1}^{d_s} \Lambda_i P_{e,i}^{(l)} \\
&= \sum_{i=1}^{d_s} \Lambda_i \int_{-\infty}^{0} \frac{1}{\sqrt{4\pi\mu_i^{(l)}}} e^{-\left(t-\mu_i^{(l)}\right)^2 / 4\mu_i^{(l)}} dt \\
&= \sum_{i=1}^{d_s} \Lambda_i Q\left(\sqrt{\mu_i^{(l)}/2}\right)
\end{aligned} \qquad (14)$$

where $Q(x)$ is the tail probability of a standard normal distribution.

*B. Optimization of degree distribution*

The performance of SLT codes mainly depends on check node degree distribution $\Omega(x)$ so in this part we will find a way to optimize the distribution. Recall that all messages passing from source nodes to check nodes in some round $l$ are i.i.d. Gaussian with mean $\mu_Q^{(l)}$. During BP decoding, the error probability will decrease from iteration to iteration when the condition in (15) satisfies.

$$\mu_Q^{(l+1)} > \mu_Q^{(l)} \qquad (15)$$

By $\alpha$ and $\beta$ we denote the average degrees of source nodes and check nodes, respectively. Poisson distribution $\Lambda(x)$ is specified by its mean $\alpha$. By using the updating rules we have achieved with GA, (15) can be expanded as [6]

$$\alpha \sum_{j=1}^{d_c} \omega_j f_j\left(\mu_Q^{(l)}\right) + \frac{2}{\sigma_n^2} > \mu_Q^{(l)} \qquad (16)$$

where

$$f_j\left(\mu_Q^{(l)}\right) = \varphi^{-1}\left(1 - \left(1 - \varphi\left(\frac{2}{\sigma_n^2}\right)\right)\left(1 - \varphi\left(\mu_Q^{(l)}\right)\right)^{j-1}\right)$$

Note that (16) is a linear inequality with unknown coefficients of check node degree distribution for all $\mu_Q^{(l)} > 0$. Actually it is not necessary for all positive $\mu_Q^{(l)}$ to hold and we just need to assume (16) holds for a range of $\mu_Q^{(l)}$, say $\mu_Q^{(l)} \in (0, \mu_0]$.

Meanwhile, to achieve a certain error probability, we may want $\varepsilon$ to be as small as possible. As $\sum_{j=1}^{d_c} \omega_j / j = \alpha / \beta = \varepsilon$, our objective is to minimize $\sum_{j=1}^{d_c} \omega_j / j$. In practice, we can use a linear programming (LP) to solve the problem. As in the LP procedure (17), we fix $d_c$, $\sigma_n^2$, $\mu_0$ and integer $L$ in advance. $\mu_k$ $(k = 0, \ldots, L-1)$ are $L$ equidistant points in the interval $(0, \mu_0]$.

$$\min \ \alpha \sum_{j=1}^{d_c} \frac{\omega_j}{j}$$
$$\text{s.t.} \ \alpha \sum_{j=1}^{d_c} \omega_j f_j(\mu_k) + \frac{2}{\sigma_n^2} > \mu_k \quad \forall k = 0, \ldots, L-1, \ \mu_k \in (0, \mu_0] \quad (17)$$
$$\sum_{j=1}^{d_c} \omega_j = 1$$
$$\omega_j \geq 0 \quad \forall j = 1, \ldots d_c$$

Note that there is no need to fix $\alpha$ because we can treat $\alpha \omega_j$ as a whole. By doing so, the second constraint in (17) should be deleted. Once LP procedure is done, (18) can be used to achieve the check node degree distribution

$$\Omega(x) = \frac{\int_0^x \omega(z) dz}{\int_0^1 \omega(z) dz} \quad (18)$$

We include our optimization results for the value $\sigma_n^2 = 1$ in Table I. Fig. 6 gives a plot of error probability versus overhead by using those optimal distributions. We also compare our distribution with that of [13] in Fig. 7.

TABLE I. LP RESULTS FOR DIFFERENT MAXIMUM DEGREES

| Maximum degree $d_c$ | Check node degree distribution $\Omega(x)$ |
|---|---|
| 20 | $\Omega_1(x) = 0.7361x^5 + 0.2639x^{20}$ |
| 50 | $\Omega_2(x) = 0.3189x^5 + 0.5713x^6 + 0.1098x^{50}$ |
| 100 | $\Omega_3(x) = 0.8966x^6 + 0.0333x^{34} + 0.0701x^{100}$ |

IV. LOWER BOUND ANALYSIS

In this section we will carry out some lower bound analyses to investigate the performance of SLT codes over AWGN channel.

*Lemma 1:* For a source node of degree $i$, when overhead $\varepsilon$ goes to infinity, LLR messages are Gaussian $\mathcal{N}\left(2(i+1)/\sigma_n^2, 4(i+1)/\sigma_n^2\right)$.

*Proof:* As we mentioned in (15), mean of LLR messages of variable nodes must increase under iterations so a successful decoding progress can be assured. And $\tanh(x)$ is close to 1 when $x$ is large enough, so we can assume that LLR messages passing from source nodes to check nodes are perfect [10] in the last iteration of BP decoding. Under such assumption, (5) can be simplified as

$$R^{(l)} = 2\tanh^{-1}\left[\tanh\left(\frac{Z}{2}\right)\right] \quad (19)$$

which means $R$ shares the same distribution with $Z$. Recall that LLR messages of the channel are Gaussian $\mathcal{N}\left(2/\sigma_n^2, 4/\sigma_n^2\right)$, so $R \sim \mathcal{N}\left(2/\sigma_n^2, 4/\sigma_n^2\right)$. Furthermore, LLR messages passing from check nodes to source nodes in (6) are sum of i.i.d Gaussian variables. Specifically, for source nodes of degree $i$, messages are Gaussian $\mathcal{N}\left(2(i+1)/\sigma_n^2, 4(i+1)/\sigma_n^2\right)$ after iterations.

□

In previous section, we have gained BER expression by using GA. Furthermore, the lower bound of SLT codes can be achieved by enforcing *Lemma 1* into (14), i.e.,

$$P_e \geq \sum_{i=1}^{d_s} \Lambda_i Q\left(\frac{\sqrt{i+1}}{\sigma_n}\right) \tag{20}$$

here we define

$$\text{LB}_1 = \sum_{i=1}^{d_s} \Lambda_i Q\left(\frac{\sqrt{i+1}}{\sigma_n}\right) \tag{21}$$

We demonstrate GA asymptotic performance and lower bound $\text{LB}_1$ in Fig. 5. Note that base-10 logarithm of asymptotic performance in Fig. 5 is approximately linear with respect to overhead $\varepsilon$ when $\varepsilon$ is large enough. So in the following, we will give another form of lower bound.

***Theorem 1:*** When BPSK modulated SLT codes are transmitted through AWGN channel, the lower bound of BER performance can be approximately expressed as

$$\text{LB}_2 \square \frac{1}{12} g(\sigma_n) \cdot e^{(g(\sigma_n)-1)\beta\varepsilon} \tag{22}$$

where

$$g(\sigma_n) = e^{-\frac{1}{2\sigma_n^2}}.$$

***Proof:*** By taking base-10 logarithm of inequality in (20) and enforcing Poisson distribution $\Lambda(x)$ we get

$$\lg P_e \geq -\beta\varepsilon \lg e + \lg\left(\sum_{i=1}^{+\infty} \frac{(\beta\varepsilon)^i}{i!} Q\left(\frac{\sqrt{i+1}}{\sigma_n}\right)\right) \tag{23}$$

Tail probability of a standard normal distribution $Q(x)$ can be approximately expressed [22] as

$$Q(x) \approx \frac{1}{12} e^{-\frac{x^2}{2}} + \frac{1}{4} e^{-\frac{2x^2}{3}} \tag{24}$$

and actually (24) is the upper bound of $Q(x)$. Note that there are two terms of approximate $Q(x)$ and in the following we have to just keep only one of them to obtain a linear relationship between base-10 logarithm of BER and overhead. Let $Q_1(x)$ and $Q_2(x)$ denote the first and the second term, respectively. Namely,

$$Q_1(x) = \frac{1}{12} e^{-\frac{x^2}{2}}$$

$$Q_2(x) = \frac{1}{4} e^{-\frac{2x^2}{3}}$$

and they are both depicted with respect to degrees in Fig. 3. It is implied that $Q_1(x)$ is closer to $Q(x)$ than $Q_2(x)$ for large degrees. Consider the product of Poisson distribution and Q-function in (20) and we demonstrate it in Fig. 4. Clearly, $Q_1(x)$ is much more accurate when calculating BER. So $Q(x)$ is approximately expressed as (25) in this paper

$$Q(x) \approx \frac{1}{12} e^{-\frac{x^2}{2}} \tag{25}$$

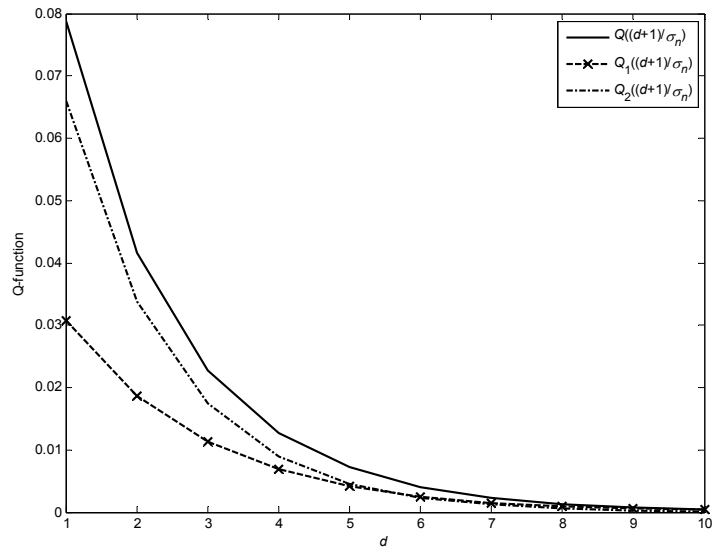

Fig. 3. Comparison between $Q_1(x)$ and $Q_2(x)$ with respect to degree $d$, where $\sigma_n$ is set to 1.

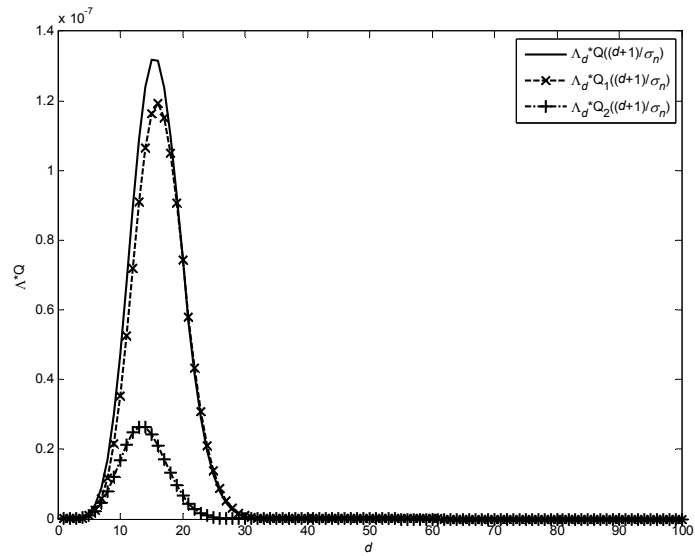

Fig. 4. Comparison between product of Poisson distribution and different Q-functions, where $\sigma_n$ is set to 1.

By using (25), (23) can be expanded as

$$\begin{aligned}
\lg P_e &\geq -\beta\varepsilon \lg e + \lg\left(\sum \frac{(\beta\varepsilon)^i}{i!} Q\left(\frac{\sqrt{i+1}}{\sigma_n}\right)\right) \\
&\approx -\beta\varepsilon \lg e + \lg\left(\sum \frac{(\beta\varepsilon)^i}{i!} \cdot \frac{1}{12} e^{-\frac{i+1}{2\sigma_n^2}}\right) \\
&= -\beta\varepsilon \lg e + \lg\left(\frac{1}{12} e^{-\frac{1}{2\sigma_n^2}} \sum \frac{(\beta\varepsilon)^i}{i!} \left(e^{-\frac{1}{2\sigma_n^2}}\right)^i\right) \\
&= -\beta\varepsilon \lg e + \lg\left(\frac{1}{12} e^{-\frac{1}{2\sigma_n^2}} \left(e^{\beta\varepsilon e^{-\frac{1}{2\sigma_n^2}}}\right)\right) \\
&= \left(e^{-\frac{1}{2\sigma_n^2}} - 1\right)\beta\varepsilon \lg e + \lg\left(\frac{1}{12} e^{-\frac{1}{2\sigma_n^2}}\right)
\end{aligned} \qquad (26)$$

Obviously, a new approximate lower bound can be defined as

$$\begin{aligned}
\text{LB}_2 &= 10^{\left(e^{-\frac{1}{2\sigma_n^2}}-1\right)\beta\varepsilon \lg e + \lg\left(\frac{1}{12}e^{-\frac{1}{2\sigma_n^2}}\right)} \\
&= \frac{1}{12} g(\sigma_n) \cdot e^{(g(\sigma_n)-1)\beta\varepsilon}
\end{aligned} \qquad (27)$$

where

$$g(\sigma_n) = e^{-\frac{1}{2\sigma_n^2}}$$

And we are done with the proof. □

The logarithm of our new lower bound LB$_2$ is a linear function of $\varepsilon$, which verifies the previous observation. LB$_2$ is depicted in Fig. 5 as well.

## V. SIMULATION RESULTS

In this section, we offer some simulation results of our work. Throughout the whole simulations, the channel is AWGN with $\sigma_n^2 = 1$, which means signal-to-noise-ratio (SNR) is 0dB under BPSK modulation.

First different lower bounds LB$_1$ and LB$_2$ are plotted in Fig. 5 as well as asymptotic performance of GA. $\Omega_3(x)$ is chosen as the check node degree distribution. The result shows that these three curves are close to each other gradually as $\varepsilon$ increases and it verifies our lower bound analyses.

Then we demonstrate BER performance of our optimal check node degree distributions in Fig. 6, where $\Omega_1(x)$ and $\Omega_3(x)$ are chosen. In order to study the impact of the length of SLT codes, we set $K = 1000$, 2000, 4000 for different distributions. Clearly, the length is infinity when using GA. As can be seen, for a certain distribution, BER performance gets better and better when $K$ increases to infinity. And also, performance of GA is the ideal extreme for finite length. Moreover, SLT codes of large average check node degree can outperform those of small average degree, which is clear in Fig. 6.

At last, we compare our optimal distribution with that of [13]. The result is shown in Fig. 7. To be fair, criterion of comparison is the same average check node degree $\beta$. We choose $\Omega_3(x)$ with $\beta \approx 14$, which is almost the same as the distribution in [13] with parameters $c = 0.3$, $\delta = 0.5$. Simulations with $K = 2000$ and GA indicate that our optimal distribution can offer better BER performance. Besides, lower bounds of those two distributions tend to be the same. This can be explained using lower bound expression in (27) due to the same $\sigma_n^2$ and $\beta$.

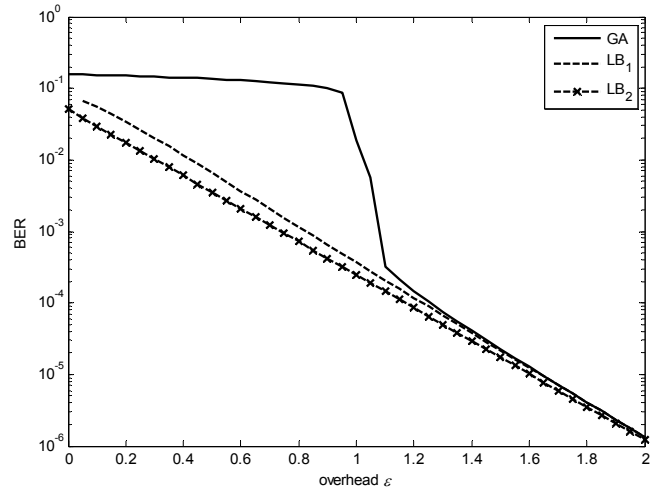

Fig. 5. Low bounds of SLT codes over AWGN channel.

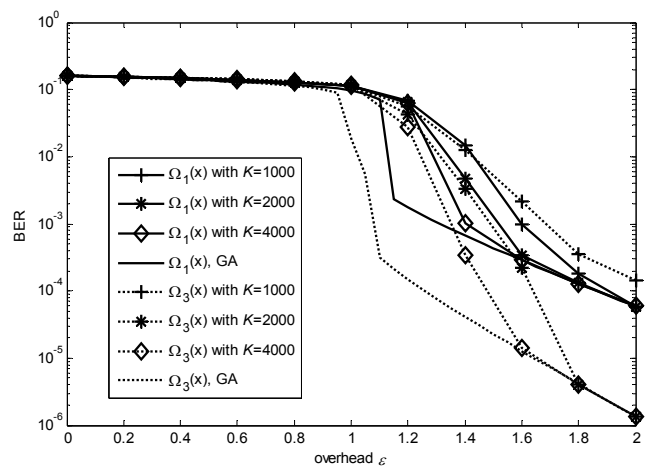

Fig. 6. BER performance of SLT codes over AWGN channel. Optimal distributions are used with different source length.

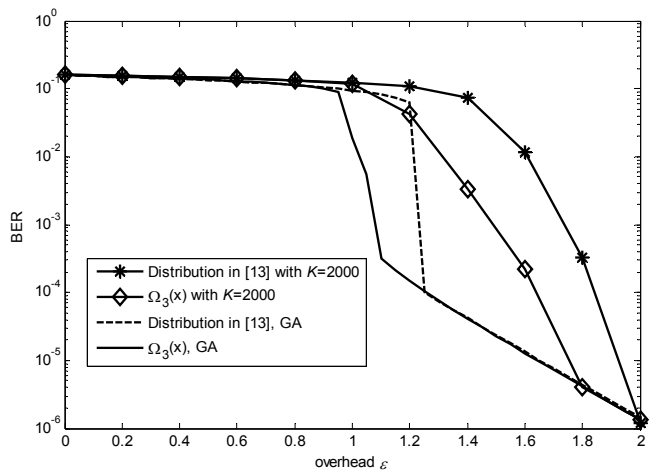

Fig. 7. Optimal distribution $\Omega_3(x)$ are compared with that of [13], with $K = 2000$ as well as Gaussian approximation.

## VI. CONCLUSION

In this paper we fully demonstrate the encoding and decoding algorithms of SLT codes first. SLT codes are the systematic form of LT codes, which consist of input symbols and encoded LT symbols. Afterwards, bipartite graph for BP decoding is provided. We simplify the bipartite graph by just keeping source nodes and check nodes on opposite sides and give the updating rules of LLR messages. Then we use GA to analyze asymptotic error probability of iterative decoding. An LP programming is proposed to optimize check node degree distribution and results are provided. Lower bound of SLT codes are also studied and we propose two low bound expressions. Finally, we do some simulations to validate our work. Results of finite length SLT codes verify the asymptotic performance. Moreover, our optimal distribution outperforms the distribution in [13].

## ACKNOWLEDGMENT

The work was supported by the National Natural Science Foundation of China (grant. No 6147